\documentclass[runningheads]{llncs}

\usepackage[T1]{fontenc}

\usepackage[T1]{fontenc}
\usepackage{graphicx}
\usepackage{multirow}
\usepackage{balance}
\usepackage{stfloats}
\usepackage{xcolor}
\usepackage{float}
\usepackage{url}
\usepackage{longtable}

\usepackage[table]{xcolor}
\usepackage{tikz}
\usepackage{diagbox}
\usetikzlibrary{positioning,fit,shapes}

\usepackage{tablefootnote}
\usepackage{listings}
\usepackage[edges]{forest}
\usepackage{tcolorbox}
\usepackage{mdframed}
\usepackage{fontawesome5}

\definecolor{lightblue}{RGB}{0,0,100}
\definecolor{purplish}{HTML}{D8DFE3}
\definecolor{purplishlight}{HTML}{EBEFF3}
\definecolor{purplishdark}{HTML}{007ACC}
\definecolor{myorange}{HTML}{007ACC}
\definecolor{lightpurple}{RGB}{200,229,204}
\definecolor{darkpurple}{RGB}{200,140,0}

\newmdenv[
  backgroundcolor=blue!9,
  leftline=true,
  rightline=false,
  topline=false,
  bottomline=false,
  linecolor=black,
  linewidth=3pt,
  innerleftmargin=10pt,
  innerrightmargin=10pt,
  innertopmargin=8pt,
  innerbottommargin=8pt,
  skipabove=8pt,
  skipbelow=8pt
]{rqbox}

\newmdenv[
  backgroundcolor=green!8,
  leftline=true,
  rightline=false,
  topline=false,
  bottomline=false,
  linecolor=black,
  linewidth=3pt,
  innerleftmargin=10pt,
  innerrightmargin=10pt,
  innertopmargin=8pt,
  innerbottommargin=8pt,
  skipabove=8pt,
  skipbelow=8pt
]{resultbox}

\newtcolorbox{MyBox}{
  colback=white,
  colframe=lightblue,
  fonttitle=\bfseries,
  coltitle=black,
  sharp corners,
  boxrule=1pt,
  left=5pt,
  right=5pt,
  top=5pt,
  bottom=5pt,
  breakable
}

\newtcolorbox[auto counter,number within=section]{promptbox}[2]{
    nameref=#1,
    title=\small{#1},
    enhanced,
    attach boxed title to top left={yshift=-6pt,xshift=8pt},
    boxed title style={size=small,boxsep=1pt},
    colframe=myorange,
    colback=white,
    colbacktitle=myorange,
    boxsep=2pt,
    left=2pt,
    right=2pt,
    top=6pt,
    bottom=2pt,
    middle=2pt
}

\newmdenv[
  backgroundcolor=lightpurple,
  linecolor=darkpurple,
  linewidth=3pt,
  leftline=true,
  topline=false,
  bottomline=false,
  rightline=false,
  innerleftmargin=10pt,
  innerrightmargin=10pt,
  innertopmargin=5pt,
  innerbottommargin=5pt,
  skipabove=10pt,
  skipbelow=10pt
]{participantbox}

\begin{document}

\title{How Software Engineering Students Use LLMs to Write Research Papers: An Experience Report}

\titlerunning{How Software Engineering Students Use LLMs to Write Research Papers}

\author{
Ronnie de Souza Santos\inst{1,2}
\and
Maria Teresa Baldassarre\inst{3}
\and
Cleyton Magalhães\inst{4}
\and
Italo Santos\inst{5}
}

\authorrunning{R. de Souza Santos et al.}

\institute{
University of Calgary, Canada
\and
CESAR School, Brazil
\and
University of Bari, Italy
\and
Universidade Federal Rural de Pernambuco (UFRPE), Brazil
\and
University of Hawai‘i at Mānoa, USA
\\
\email{
ronniedesouzasantos@ucalgary.ca,
ress@cesar.school,
mariateresa.baldassarre@uniba.it,
cleyton.vanut@ufrpe.br,
isantos3@hawaii.edu}
}

\maketitle

\begin{abstract}

Large language models are now part of software engineering education, including activities involving empirical software engineering and evidence synthesis. This paper reports an educational experience involving the integration of reflective LLM use into an empirical methods assignment in a third-year software architecture course. Students were asked to develop a short research paper using either a rapid review or a gray literature review methodology and to disclose how LLMs were used throughout the assignment. We analyzed 146 student disclosure statements using a cross-analysis process combining LLM-assisted categorization with manual verification and refinement by the researchers. The reflections describe how students incorporated LLMs during activities such as brainstorming, methodological clarification, organization of findings, and writing refinement, while also reporting concerns regarding inaccuracies and verification of generated content. This experience report discusses lessons learned and educational implications for integrating AI-assisted technologies into empirical software engineering education.


\end{abstract}

\section{Introduction}
\label{sec:introduction}

Large language models (LLMs) are increasingly present in educational settings, supporting activities such as tutoring, feedback generation, content creation, and writing assistance~\cite{moore2023empowering,wen2024ai,wiktor2024ai}. In software engineering education, these tools are now part of students’ everyday academic workflows, including programming activities, brainstorming, information synthesis, and academic writing~\cite{filippi2024large,arora2025analyzing,liu2024integrating}. At the same time, their growing adoption has raised questions regarding authorship, reliability, critical thinking, academic integrity, and the role of AI in learning processes~\cite{yan2024practical,grande2024student,razafinirina2024pedagogical,pelaez2024impact}.

Current discussions in software engineering education increasingly argue that LLMs should not be treated solely as coding assistants or productivity tools, but rather as technologies that require critical and reflective engagement~\cite{kirova2024software,pereira2024leveraging,filippi2024large}. In this context, educators are beginning to rethink how assignments are designed, how students are encouraged to interact with AI systems, and how responsible use can be incorporated into teaching practices. These discussions become particularly relevant in empirical software engineering education, where students are expected to develop skills related to evidence synthesis, methodological reasoning, critical analysis, and communication~\cite{host2002introducing,wohlin2007empirical,pizard2021training,serebrenik2024teaching}.

This paper reports our experience integrating reflective and transparent LLM use into an empirical methods assignment in a third-year software architecture course. Students were allowed to use LLMs throughout the process of developing a short research paper, provided that they disclosed how the tools were used. The activity aimed not only to support students during the writing process but also to encourage reflection on the role of AI tools in academic work. To better understand how students interacted with LLMs in this educational setting, we analyzed 146 disclosure statements written after completion of the assignment. These statements described how students used LLMs, which aspects of the writing process were supported by the tools, and which benefits and limitations they perceived. Rather than evaluating the effectiveness of LLMs, our goal was to understand how students appropriated these technologies during an empirical writing activity and what this experience suggests for software engineering education. The observations presented in this paper provide practical insights for educators interested in incorporating LLMs into empirical software engineering assignments. In particular, our experience suggests the importance of transparency, critical verification, and explicit guidance when integrating AI-assisted technologies into academic work.

From this introduction, this paper is organized as follows. Section~\ref{sec:background} discusses empirical methods teaching in software engineering education. Section~\ref{sec:method} presents the educational context and the assignment structure. Section~\ref{sec:findings} reports the main observations from student disclosures. Section~\ref{sec:discussion} discusses lessons learned and implications for education. Finally, Section~\ref{sec:conclusion} concludes the paper.

\section{Teaching Empirical Methods} 
\label{sec:background}

Teaching empirical methods in software engineering is widely recognized as important for helping students understand how to evaluate technologies and practices systematically and using evidence rather than intuition alone~\cite{host2002introducing,wohlin2007empirical}. As empirical research has become an established component of software engineering, students are increasingly expected to understand how to design studies, collect and analyze data, and interpret findings in order to support decision making~\cite{host2002introducing,pizard2021training}. Methods such as controlled experiments, case studies, surveys, and evidence synthesis approaches provide opportunities for students to develop these competencies while strengthening their understanding of software engineering practice~\cite{pizard2021training}. Beyond methodological knowledge, empirical methods education seeks to cultivate an evidence based mindset in future software engineers. Students are expected not only to understand how empirical studies are conducted, but also how empirical evidence can be used to evaluate technologies, compare alternatives, justify decisions, and critically assess claims reported in both research and practice~\cite{wohlin2007empirical,pizard2021training}. This objective has become increasingly relevant as software engineering continues to evolve rapidly and practitioners are frequently required to assess new tools, development approaches, and emerging technologies.

To support these learning objectives, educators have adopted a variety of pedagogical approaches, including dedicated empirical software engineering courses, the integration of empirical assignments into broader software engineering curricula, and project based activities that expose students to different stages of the research process~\cite{wohlin2007empirical,serebrenik2024teaching}. Such activities commonly involve study design, data analysis, interpretation of findings, and reflection on threats to validity~\cite{host2002introducing,pizard2021training}. Through these experiences, students engage with both the theoretical and practical dimensions of empirical research~\cite{serebrenik2024teaching}. A recurring recommendation in the literature is that empirical methods should be taught through authentic experiences rather than through lectures alone~\cite{host2002introducing,wohlin2007empirical,pizard2021training}. Students benefit from participating in the complete empirical cycle, including identifying research questions, selecting appropriate methods, collecting and analyzing data, interpreting results, and communicating findings. Project based assignments allow students to experience the challenges associated with conducting empirical investigations while developing practical skills that are difficult to acquire through theoretical instruction alone~\cite{host2002introducing,pizard2021training}.

Another important aspect concerns the development of critical thinking and reflection skills. Empirical software engineering requires students to evaluate the quality of evidence, identify limitations in study designs, recognize threats to validity, and understand the contextual nature of software engineering findings~\cite{wohlin2007empirical,serebrenik2024teaching}. Consequently, educational activities often encourage students to critically examine published studies, compare alternative methodological approaches, and reflect on the implications of empirical results for practice~\cite{pizard2021training}. Recent discussions have also emphasized the importance of active learning approaches when teaching empirical software engineering~\cite{serebrenik2024teaching}. Rather than positioning students as passive recipients of methodological concepts, these approaches encourage participation through discussions, collaborative activities, peer learning, and practical exercises. Such strategies provide opportunities for students to connect methodological concepts with realistic software engineering problems and support a deeper understanding of empirical reasoning.

Despite their educational value, empirical methods courses present several challenges. Students often enter these courses with varying levels of methodological and software engineering knowledge, requiring instructors to balance conceptual foundations with practical application~\cite{serebrenik2024teaching}. In addition, educators must design activities that promote learning while maintaining realistic expectations regarding research outcomes and project scope~\cite{wohlin2007empirical}. Balancing methodological rigor with accessibility can be particularly challenging when students have limited prior experience with research methods, statistics, or academic writing~\cite{pizard2021training}. Nevertheless, empirical activities have generally been reported as valuable learning experiences and remain an important component of software engineering education~\cite{host2002introducing,pizard2021training}. By combining methodological instruction with hands on investigation and critical reflection, empirical methods courses can help students develop the skills needed to evaluate evidence, reason about software engineering practices, and make informed decisions throughout their professional careers.

\section{Educational Context and Experience Report} 
\label{sec:method}

This paper reports our experience incorporating reflective and transparent LLM use into an empirical methods assignment in a third-year undergraduate software architecture course. Rather than evaluating the effectiveness of LLMs themselves, our goal was to better understand how students appropriated these tools while developing academic work and what this experience suggests for software engineering education.

\subsection{Educational Context}

The experience was conducted in a third-year undergraduate software architecture course within a software engineering program. As part of the course activities, students were asked to write a short 2--3-page paper investigating a topic related to software architecture or software design. Students were required to define a research question and apply either a Rapid Review or a Grey Literature Review methodology following structured instructional guidelines provided during the course. The assignment was designed to expose students to forms of evidence that software engineers commonly encounter when investigating technologies, practices, and design decisions. Rather than conducting original empirical studies, students worked with existing evidence sources and were asked to identify patterns, compare perspectives, and construct evidence-informed conclusions. This approach allowed students to experience key stages of empirical inquiry while keeping the activity feasible within the scope of a regular undergraduate course.

Students could choose between a Rapid Review and a Grey Literature Review depending on their interests and the nature of their research question. The Rapid Review option exposed students to peer-reviewed research and encouraged engagement with how software engineering knowledge is reported, evaluated, and synthesized. The Grey Literature Review option focused on practitioner discussions and enabled students to investigate industry experiences, challenges, and opinions that are often not represented in academic publications. Providing both options allowed students to explore different forms of evidence while recognizing that software engineering knowledge is produced by both researchers and practitioners. For the Rapid Review, students were instructed to systematically identify and analyze at least 15 peer-reviewed studies published in reputable software engineering venues within the last 10 years. In the Grey Literature Review, students were asked to analyze at least 30 practitioner-oriented online posts from sources such as StackExchange, Quora, and Dev.to. In both activities, students were expected to synthesize findings, organize evidence into categories or themes, and communicate their results using an academic structure. The assignment therefore required students to move beyond summarization by identifying recurring observations, contrasting viewpoints, and broader implications emerging from the collected evidence.

Students were explicitly allowed to use LLMs throughout the assignment development process. This included support for brainstorming, refining research questions, improving writing clarity, organizing findings, and understanding methodological concepts. However, students were instructed that the papers should reflect their own reasoning and synthesis rather than fully AI-generated content. To encourage transparency and reflection, students were required to include a short disclosure statement describing whether and how LLMs were used during the assignment. These disclosures were not graded and were included as part of broader course guidelines regarding AI use across assignments.

\subsection{Reflection Collection}

The disclosure statements were collected only after the course had concluded and final grades had been released. This procedure ensured that no academic consequences could be associated with the disclosures and that students could reflect on their experiences without affecting evaluation outcomes. To preserve anonymity, two Teaching Assistants extracted the disclosure statements from the submitted assignments, removed identifying information, and organized the material into a shared spreadsheet. The statements were inserted in randomized order to reduce the possibility of associating disclosures with specific students. In total, 146 anonymized statements were compiled for reflection and analysis.

\subsection{Cross Analysis of Student Reflections}

To characterize how students described their use of LLMs, we conducted a cross-analysis combining LLM-assisted categorization with manual verification and refinement by the researchers. The process was intentionally designed as a reflective educational analysis rather than a fully automated classification process. The analysis was conducted in four phases. \\

\textit{Phase 1: Familiarization with the Reflections.}  
Initially, the researchers independently read the full set of 146 disclosure statements multiple times to become familiar with the reflections and identify recurring patterns related to how students used LLMs during the assignment. During this stage, attention was given to recurring descriptions of writing support, methodological assistance, brainstorming activities, perceived benefits, and reported challenges. \\

\textit{Phase 2: Initial LLM-Assisted Categorization.}  
To support the organization of recurring patterns across the disclosures, ChatGPT-4.0 was used to assist with excerpt identification and preliminary categorization tasks. The prompts focused on dimensions commonly emphasized in reviewer guidelines from major software engineering venues such as ICSE, ICSME, ESEM, EASE, and the \textit{Empirical Software Engineering} journal. These venues frequently instruct reviewers to evaluate aspects such as novelty, rigor, relevance, transparency, and presentation quality when assessing research papers.

Based on these recurring dimensions, the analysis was organized around the following categories:

\begin{itemize}
    \item \textbf{Novelty}: uses related to brainstorming, topic exploration, research question definition, and understanding unfamiliar concepts;

    \item \textbf{Rigor}: uses associated with methodological clarification, organization of research procedures, identification of evidence sources, and support for empirical research activities;

    \item \textbf{Relevance}: uses focused on improving the communication, interpretation, coherence, and significance of findings and conclusions;

    \item \textbf{Transparency}: uses related to improving explanations of research procedures, clarity of reporting, traceability of information, and communication of how findings were obtained or analyzed;

    \item \textbf{Presentation}: uses associated with grammar correction, writing refinement, formatting, readability, sentence structure, and overall presentation quality.
\end{itemize}

ChatGPT-4.0 was used to identify excerpts potentially associated with these categories and to flag explicit mentions of perceived benefits and challenges related to LLM use. An illustrative example of the prompt structure used during the categorization process is presented below.

\begin{resultbox}
\faChartBar\ \textbf{Illustrative Analysis Prompt}\\

\textbf{Instruction}: Analyze the following student disclosure statements and identify excerpts where students explicitly describe how LLMs supported academic writing activities. Categorize the excerpts according to the following dimensions commonly considered in software engineering peer review: Novelty, Rigor, Relevance, Transparency, and Presentation.

\textbf{Guidelines}: Only classify excerpts when the student explicitly describes the use or perceived impact of the tool. Do not infer meanings that are not directly stated in the disclosure statements.

\textbf{Output}: Return the identified category together with the corresponding excerpt from the disclosure statement.
\end{resultbox}

\textit{Phase 3: Manual Verification and Refinement.}  
After the initial AI-assisted categorization, the researchers manually reviewed all classifications and extracted quotations. During this verification process, the researchers checked whether the excerpts accurately reflected the assigned category, removed ambiguous classifications, adjusted labels when necessary, and refined the interpretation of the disclosures to preserve contextual meaning.

Particular attention was given to avoiding overinterpretation. Only disclosures in which students explicitly described a use, challenge, or perception related to LLMs were retained in the final categorization. Statements considered vague, indirect, or unsupported were excluded from the analysis. The researchers also compared classifications collaboratively and discussed disagreements until consensus was reached. \\

\textit{Phase 4: Consolidation of Educational Observations.}  
Finally, the refined categories and excerpts were consolidated into broader educational observations regarding how students integrated LLMs into empirical writing activities. At this stage, the researchers focused on recurring educational patterns, including how students perceived LLMs as writing assistants, methodological aids, brainstorming partners, and sources of concern related to hallucinations, inaccuracies, and meaning distortion. This iterative interaction between AI-assisted support and researcher interpretation allowed the organization of recurring educational observations while preserving the contextual richness of the student reflections.

\subsection{Ethics}
All procedures followed institutional ethical guidelines for research involving human participants. Disclosure statements were analyzed only after final grades had been released, ensuring that participation had no influence on academic evaluation. The research team had access only to anonymized disclosures and did not interact with identifiable student data during the analysis process.

Due to the nature of the disclosures and the educational setting, the complete dataset cannot be publicly shared. However, anonymized excerpts are included throughout the paper to illustrate the observations discussed.
\section{Educational Observations}
\label{sec:findings}

Across the 146 student disclosure statements collected after the assignment, several recurring patterns emerged regarding how students incorporated LLMs into their academic writing activities. Most students described using LLMs to support presentation and writing quality, while smaller groups reported using the tools for idea generation, methodological support, articulation of findings, and transparency-related tasks. A small number of students also explicitly stated that they chose not to use LLMs during the assignment.

Most students reported using ChatGPT as their primary LLM during the assignment. A smaller number mentioned other tools, including Claude, Gemini, GitHub Copilot, Bing AI, and Deepseek. In some cases, students described combining multiple LLMs depending on the activity being performed, particularly for brainstorming, writing refinement, or summarization tasks. These observations suggest that while ChatGPT dominated usage patterns, students were also experimenting with different AI tools throughout the assignment process.

\small

\begin{longtable}{p{2.5cm}|p{8.5cm}}
\caption{Illustrative Student Disclosures by Type of LLM Use}
\label{tab:llmusequotes}\\
\hline
\textbf{Category} & \textbf{Illustrative Quotes}\\
\hline \hline
\endfirsthead

\hline
\textbf{Category} & \textbf{Illustrative Quotes}\\
\hline
\endhead

\textbf{Presentation}
&
\textit{“This paper leveraged ChatGPT and Google Gemini for the improvement of sentence structure, spelling, grammar, and Latex formatting.”} (D001)

\smallskip

\textit{“ChatGPT, a generative AI tool, was used in the writing process of this paper for grammatical checks and proper sentence phrasing.”} (D005)

\smallskip

\textit{“In this paper, ChatGPT and Grammarly AI were tools used for grammar, spelling/thesaurus, checking if content was in accordance with IEEE standard.”} (D006)
\\
\hline \hline

\textbf{Novelty}
&
\textit{“In the early brainstorming stages of this paper, ChatGPT was used to gain a broad understanding (...)”} (D013)

\smallskip

\textit{“The author of this paper used ChatGPT-4o to generate ideas for the topic of this paper.”} (D053)

\smallskip

\textit{“I used ChatGPT to understand the key concepts of the papers.”} (D120)
\\
\hline \hline

\textbf{Relevance}
&
\textit{“(...) it helped me word and phrase my findings and thoughts in a more cohesive and clear manner.”} (D016)

\smallskip

\textit{“It helped me break down complicated ideas, explain difficult terms, and ensure that my explanations made sense.”} (D111)

\smallskip

\textit{“In this paper, ChatGPT was used to (...) refining clarity in written explanations.”} (D134)
\\
\hline \hline

\textbf{Rigor}
&
\textit{“(...) to create a step-by-step guide for conducting research using the prompt ‘can you devise a plan to research and write the paper based on that question?’”} (D024)

\smallskip

\textit{“In this paper, the DeepSeekR1 LLM served as a research partner, providing a second perspective during the study screening process by evaluating inclusion criteria that I originally created.”} (D099)

\smallskip

\textit{“Furthermore, I used it to give me key bullet points from some of the papers I researched.”} (D133)
\\
\hline \hline

\textbf{Transparency}
&
\textit{“A blueprint of the structure of this paper was provided, to which the tool identified areas where additional material could enhance the explanation.”} (D083)

\smallskip

\textit{“In this paper, ChatGPT was used to enhance the reader’s comprehension (...) making the paper more transparent and effective.”} (D093)
\\
\hline \hline

\end{longtable}

\subsection{LLMs as Writing and Research Support Tools}

The most frequently reported use of LLMs involved \textbf{Presentation}. Students commonly described using the tools to revise grammar, improve sentence structure, increase clarity, and adapt the tone of their papers to a more academic style. Several students characterized the LLM as a “final editor” used before submission to polish the text and improve readability. Students also reported using LLMs to support citation formatting, particularly in BibTeX and \LaTeX{}, and to help reorganize sections to better align with the assignment template. These observations suggest that students frequently relied on LLMs to navigate academic writing conventions and improve the overall presentation of their work.

Another recurring pattern involved the use of LLMs for \textbf{Novelty}. Students described using the tools to explore possible research topics, narrow broad themes, identify potential research questions, and better understand unfamiliar concepts related to software architecture and software design. Some students reported iterating over multiple prompts to compare possible directions for their papers or to better define the scope of their investigations. In several disclosures, students described the tools as interactive brainstorming partners that helped initiate early-stage academic reasoning and topic exploration.

A smaller group of students described using LLMs to support \textbf{Relevance}. In these cases, students reported asking the tools for assistance in explaining the relevance of their observations, synthesizing conclusions, or improving the communication of implications identified during the assignment. Rather than generating findings directly, the tools were often used to help students organize and phrase their interpretations more clearly. These reflections suggest that some students perceived LLMs as support mechanisms for strengthening coherence and improving how results and implications were communicated in their papers.

Some students also described using LLMs to support \textbf{Rigor}. These uses included asking for clarification about rapid reviews and gray literature reviews, obtaining suggestions for structuring research procedures, and improving descriptions of data organization and analysis activities. A few students also reported using the tools to help identify possible evidence sources and organize categories or themes across collected materials. From an educational perspective, these observations indicate that students may perceive LLMs not only as writing assistants, but also as tools capable of supporting understanding of empirical software engineering practices and research organization activities.

Finally, a small number of students reported using LLMs to support \textbf{Transparency}. In these cases, students described using the tools to improve explanations regarding how information was collected, categorized, or analyzed. Some students also used LLMs to review whether their descriptions of research procedures were sufficiently understandable and reproducible. Although less frequent, these observations suggest that some students were already reflecting on how AI tools could support communication of research processes, methodological clarity, and transparency during academic writing activities.

\subsection{Student Reflections on Benefits and Challenges}

Students reported a wide range of experiences regarding the incorporation of LLMs into academic writing activities. While some disclosures were brief and technical, others provided detailed reflections about how these tools affected writing organization, comprehension, confidence, and learning processes throughout the assignment. Many students described LLMs as useful companions during the writing process. Commonly reported benefits included improved grammar, better organization of ideas, assistance with summarization, and support for understanding complex concepts. Several students also emphasized that LLMs helped make academic writing feel more manageable, particularly when dealing with unfamiliar terminology, empirical methods, or formal academic structures.

For example, student D138 stated: \textit{``AI did help improve my writing process by helping me to avoid repeating the same ideas again and again.''} Similarly, student D109 explained: \textit{``Additionally, it provided support in understanding complex concepts and rephrasing them in simpler terms so I could grasp them better.''} These reflections suggest that many students perceived LLMs as tools capable of supporting both communication and comprehension during empirical writing activities. In particular, several disclosures associated AI support with increased clarity, confidence, organization, and reduced difficulty during the writing process.

At the same time, students also expressed concerns regarding limitations and risks associated with LLM use. Several disclosures mentioned hallucinations, inaccuracies, lack of contextual understanding, and the possibility that AI-generated suggestions could unintentionally distort the intended meaning of the text. Some students also described difficulties determining whether generated outputs were reliable or sufficiently aligned with their intended arguments. These concerns indicate that, despite the perceived usefulness of the tools, students remained aware of practical limitations and risks associated with relying on AI-generated content during academic work.

For example, participant D026 stated: \textit{``A key challenge being that the AI-assisted refinements altered the intended meaning or introduce unintended biases.''} Another student, D018, reported: \textit{``However, hallucinations were an issue as in certain cases it made up fake information in terms of certain topics.''} These disclosures suggest that students were not simply accepting generated outputs without reflection. Instead, many students described the need to verify information, critically evaluate generated suggestions, and preserve authorship and meaning during the writing process. Considering the reflections as a whole, the statements revealed a combination of perceived benefits and challenges associated with the use of LLMs in academic writing activities.
\section{Lessons Learned and Educational Implications}
\label{sec:discussion}

Prior literature on teaching empirical software engineering emphasizes exposing students to authentic research activities while developing skills related to study design, evidence evaluation, critical reasoning, and communication of findings~\cite{host2002introducing,wohlin2007empirical,pizard2021training,serebrenik2024teaching}. Existing educational approaches commonly encourage students to engage with empirical methods through project-based activities, evidence synthesis exercises, and structured reflection on methodological decisions. Our experience suggests that the growing presence of LLMs does not fundamentally alter these educational objectives. Instead, it changes the environment in which students perform these activities. Rather than acting solely as writing assistants, students frequently used LLMs as interactive companions during brainstorming, methodological clarification, organization of findings, and refinement of explanations. Based on these observations, we identified the following lessons learned:

\begin{itemize}

\item \textbf{Allow and regulate rather than prohibit.} Students integrated LLMs into multiple stages of the assignment process. Educational strategies may therefore be more effective when they focus on responsible use and accountability rather than attempting to eliminate the use of these tools altogether.

\item \textbf{Require explicit disclosure of AI use.} Asking students to document how LLMs contributed to their work encouraged reflection regarding benefits, limitations, verification practices, and decision-making. Disclosure activities can promote transparency while providing instructors with insights into emerging usage patterns.

\item \textbf{Teach verification as a core empirical skill.} Students frequently reported concerns regarding hallucinations, inaccuracies, and unintended changes in meaning. Assignments should explicitly emphasize validation of AI-generated information against original sources and empirical evidence.

\item \textbf{Design assignments that assess reasoning rather than text production alone.} While LLMs often supported writing quality and organization, students remained responsible for interpreting evidence, making methodological decisions, and justifying conclusions. Assessment strategies should place greater emphasis on these higher-order activities.

\item \textbf{Provide guidance on acceptable and unacceptable uses of LLMs.} Students reported a broad spectrum of uses, ranging from grammar correction to methodological support and idea generation. Clear expectations can help students distinguish between legitimate assistance and uses that may compromise learning objectives.

\item \textbf{Incorporate discussions about AI limitations into empirical methods training.} Concerns regarding reliability, bias, and factual correctness emerged repeatedly in the disclosures. These limitations can be used as opportunities to reinforce concepts such as validity, evidence quality, critical evaluation, and research rigor.

\item \textbf{Use LLMs as opportunities to strengthen methodological reflection.} Some students relied on LLMs to better understand empirical procedures and evidence synthesis approaches. Encouraging students to critically evaluate and justify AI-generated methodological suggestions may support deeper engagement with empirical concepts.

\end{itemize}

More broadly, we suggest that software engineering education may need to move beyond discussions centered on whether students should use LLMs and instead focus on how such tools can be incorporated without compromising learning objectives. Empirical software engineering courses are intended to develop critical thinking, evidence evaluation, methodological reasoning, and communication skills. Our findings suggest that these competencies remain essential in AI-enabled educational environments and may become even more important as students gain access to increasingly capable generative tools. Consequently, future educational practices may benefit from integrating transparency, verification, and responsible AI use as explicit learning outcomes alongside traditional empirical software engineering competencies.
\section{Conclusions}
\label{sec:conclusion}

This paper reports an educational experience involving the integration of reflective LLM use into an empirical software engineering assignment in an undergraduate software engineering course. Through the analysis of 146 student disclosure statements, we explored how students described the incorporation of LLMs while conducting activities related to rapid reviews and gray literature reviews. The analysis focused on five recurring dimensions of use: Presentation, Novelty, Relevance, Rigor, and Transparency. The reflections showed that students used LLMs not only for writing refinement activities, but also for brainstorming, methodological clarification, organization of findings, and communication of empirical results. At the same time, students frequently reported concerns involving hallucinations, inaccuracies, meaning distortion, and the need to verify generated outputs. These observations suggest that students are already incorporating AI tools into empirical software engineering workflows while negotiating issues related to authorship, reliability, and critical evaluation. For future work, we plan to conduct additional empirical studies involving different software engineering courses, instructional strategies, and evidence synthesis activities, such as systematic literature reviews and multivocal literature reviews, to better understand how students incorporate AI-assisted technologies into empirical software engineering workflows and learning processes.

\begin{credits}

\subsubsection{\discintname}
The authors have no competing interests to declare that are relevant to the content of this article.

\end{credits}

\bibliographystyle{splncs04}
\bibliography{bib}

\end{document}